\documentclass[a4paper,10pt]{article}
\usepackage{latexsym}
\usepackage{amsmath}
\usepackage{amssymb}
\usepackage{amscd}

\newlength{\minitwocolumn}
\makeatletter
\@addtoreset{equation}{section}
\makeatother

\title{\bf 
The critical
$A_{n-1}^{(1)}$ chain}

\author{T. Kojima${}^*$ and 
S. Yamasita${}^{\dagger}$ }

\date{\it ${}^*$Department of Mathematics,
College of Science and Technology, 
Nihon University, Chiyoda-ku, Tokyo 101-0062, Japan \\
${}^\dagger$ Information Technology Center, 
Daiwa Institute of Research Ltd.,\\
K$\bar{o}$t$\bar{o}$-ku, Tokyo 135-8460, Japan\\
~\\
{\rm \today}
}
\begin{document}

\maketitle
\begin{abstract}
We study the $A_{n-1}^{(1)}$ spin chain at the
critical regime $|q|=1$.
We give the free boson realizations of 
the type-I vertex operators
and their duals.
Using these free boson realizations,
we 
give 
the integral representations
for the correlation functions.
\end{abstract}

\section
{Introduction}
The one dimensional spin $A_{n-1}^{(1)}$ chain is described by
the Hamiltonian :
\begin{eqnarray}
{\cal H}=\sum_{k=-\infty}^\infty
\left\{
q\sum_{a,b=0 \atop{a>b}}^{n-1}
e_{aa}^{(k+1)}e_{bb}^{(k)}+q^{-1}
\sum_{a,b=0 \atop{a<b}}^{n-1}
e_{aa}^{(k+1)}e_{bb}^{(k)}
-\sum_{a,b=0 \atop{a\neq b}}^{n-1}
e_{ab}^{(k+1)}
e_{ba}^{(k)}
\right\},\label{Hamiltonian}
\end{eqnarray}
where $e_{ab}^{(k)}$ is a matrix unit acting on $k$ th site.
In this paper we consider the critical regime
$|q|=1$.
For the special case 
$n=2$ this model becomes the celebrated XXZ spin.
M.Jimbo,H.Konno and T.Miwa
\cite{JKM}
established the trace construction
of the correlation functions for
the XXZ chain at critical regime $|q|=1$.
In this paper we give 
the correlation functions for the higher rank generalization of
the XXZ chain at the critical regime $|q|=1$.
In order to write down
the correlation functions,
we need the 'dual' vertex operators.
This problem was absent from
the XXZ chain, because
the vertex operators are self-dual in this case.
We give the pair of the vertex operators and their duals
in this paper.

We give 
the $N$ point correlation functions 
which describe the ground-state
average $\langle {\cal O} \rangle$
of the local operator :
\begin{eqnarray}
{\cal O}=e_{\epsilon_1 \epsilon_1'}^{(1)}
\cdots e_{\epsilon_N \epsilon_N'}^{(N)}.
\end{eqnarray}
The correlation functions of
the critical $A_{n-1}^{(1)}$ chain
are described by the following systems of
difference equations :
\begin{eqnarray}
&&G^{(N)}(\beta_1, \cdots, \beta_j-\lambda i,
\cdots ,\beta_N|
\beta_{N+1},\cdots, \beta_{2N})
\nonumber
\\
&=&
R_{j,j-1}^{V^*V^*}(\beta_j-\beta_{j-1}-\lambda i)
\cdots
R_{j,1}^{V^*V^*}(\beta_j-\beta_{1}-\lambda i)
R_{j,2N}^{V^*V}(\beta_j-\beta_{2N})
\cdots
R_{j,N+1}^{V^*V}(\beta_j-\beta_{N+1})
\nonumber
\\
&\times&
R_{j,N}^{V^*V^*}(\beta_j-\beta_N)
\cdots
R_{j,j+1}^{V^*V^*}(\beta_j-\beta_{j+1})
G^{(N)}(\beta_1, \cdots, \beta_j, \cdots ,\beta_N|
\beta_{N+1},\cdots, \beta_{2N}),
\label{qKZ1}
\end{eqnarray}
and
\begin{eqnarray}
&&G^{(N)}(\beta_1, \cdots, \beta_N|
\beta_{N+1},\cdots, 
\beta_j+i\lambda, \cdots \beta_{2N})\nonumber
\\
&=&
R_{j+1,j}^{VV}(\beta_{j+1}-\beta_j+\lambda i)\cdots
R_{2N.j}^{VV}(\beta_{2N}-\beta_j+\lambda i)
R_{1,j}^{V^* V}(\beta_1-\beta_j)
\cdots
R_{N,j}^{V^* V}(\beta_N-\beta_j)\nonumber
\\
&\times&
R_{N+1,j}^{VV}(\beta_{N+1}-\beta_j)
\cdots
R_{j-1,j}^{VV}(\beta_{j-1}-\beta_j)
G^{(N)}(\beta_1, \cdots, \beta_N|
\beta_{N+1},\cdots, 
\beta_j, \cdots \beta_{2N}).
\label{qKZ2}
\end{eqnarray}
Here $R_{ij}^{VV}(\beta) \in {\rm End}(V^{\otimes 2N})$
signifies the matrix acting as $R^{VV}(\beta)$
on th $(i,j)-$th tensor components and as identity elsewhere.
Here $R^{VV}(\beta), R^{V^*V^*}(\beta)$ and
$R^{V^*V}(\beta)$ are given by
(\ref{RVV}), (\ref{RV*V*}) and (\ref{RV*V}).
\\
The correlation functions satisfy the restriction
equation :
\begin{eqnarray}
&&G^{(N)}(\beta+i(\pi-\lambda), \beta_2,\cdots, \beta_N|
\beta_{N+1},\cdots, \beta_{2N})_{\epsilon_1 \cdots \epsilon_N,
\epsilon_N' \cdots \epsilon_1'}\nonumber
\\
&=&\delta_{\epsilon_1,\epsilon_1'}
G^{(N-1)}(\beta_2,\cdots,\beta_N|
\beta_{N+1},\cdots,\beta_{2N-1})
_{\epsilon_2 \cdots \epsilon_N,
\epsilon_N' \cdots \epsilon_2'}.\label{restriction}
\end{eqnarray}
Here we have set
\begin{eqnarray}
&&G^{(N)}(\beta_1 \cdots \beta_N|\beta_{N+1}\cdots \beta_{2N})
\nonumber\\
&=&
\sum_{\epsilon_1 \cdots \epsilon_{2N}=0}^{n-1}
v_{\epsilon_1}\otimes \cdots \otimes v_{\epsilon_{2N}}
G^{(N)}(\beta_1 \cdots \beta_N|\beta_{N+1}\cdots \beta_{2N}
)_{\epsilon_1 \cdots \epsilon_{2N}}.
\end{eqnarray}
In this paper we set the deformation parameter
as
\begin{eqnarray}
q=-\exp\left(\frac{\pi i}{\xi}\right),
\end{eqnarray}
where $\xi >1$.\\ 
The ground state averages are obtaind from
the components $G^{(N)}$,
by taking $\lambda=2\pi$,
and specializing the spectral parameters :
\begin{eqnarray}
G^{(N)}(\beta+\pi i, \cdots , \beta+\pi i
|\beta, \cdots ,\beta)
_{\epsilon_1 \cdots \epsilon_{N},
\epsilon_N \cdots \epsilon_{1}
},
\end{eqnarray}

When we solve the diffence equations (\ref{qKZ1}), 
(\ref{qKZ2}), directly \cite{JM},
we have a difficulty. 
The difficulty in this approach is that
the solutions are determined only up to
arbitrary periodic functions,
so one has to single out in some way
the correct solutions which correspond
to the correlation functions.
When we construct the solutions
by the trace of the vertex operators,
the ambiguity of solutions are resolved.
In this paper we give 
the free boson realizations of the type-I vertex operators
and their duals, and give the trace construction
for the correlation functions of the critical $A_{n-1}^{(1)}$
spin.
In this connection, we should mention about
the work \cite{MT}, in which the authors
give the free boson realizations of the dual type-II
vertex operators of the $A_{n-1}^{(1)}$ Toda field theory
with imaginary coupling.

Now a few words about the organization of the paper.
In section 2 we review the model briefly.
In section 3 we give the defining relations of
the vertex operators.
In section 4 we give the free boson realizations
of the vertex operators.
In section 5 we give the proofs of the properties 
of the vertex operators.
In section 6 we give the integral representations
of the correlation functions.
In Appendix A we summarized the multi Gamma functions.
In Appendix B we summarized the normal ordering
of the basic operators.

\section
{The critical $A_{n-1}^{(1)}$ chain}
Let us set the $R$-matrix by
\begin{eqnarray}
R^{VV}(\beta)=r(\beta)\bar{R}(\beta),~~
r(\beta)=-
\frac
{S_2(i\beta|
\frac{2\pi}{n}\xi,2\pi
) S_2(-i\beta+\frac{2\pi}{n}|
\frac{2\pi}{n}\xi,2\pi
)}
{S_2(-i\beta|
\frac{2\pi}{n}\xi,2\pi
) S_2(i\beta+\frac{2\pi}{n}|
\frac{2\pi}{n}\xi,2\pi
)},
\end{eqnarray}
where $S_2(\beta|\omega_1,\omega_2)$ is the Multi-Sine function 
given in Appendix A.\\
The matrix $\bar{R}(\beta)$ is given as follows :
\begin{eqnarray}
\bar{R}(\beta)v_{k_1}\otimes v_{k_2}=
\sum_{j_1,j_2=0}^{n-1}
v_{j_1}\otimes v_{j_2}
\bar{R}(\beta)_{j_1 j_2}^{k_1 k_2},
\end{eqnarray}
where the nonzero entrise are 
\begin{eqnarray}
\bar{R}(\beta)_{jj}^{jj}&=&1,\\
\bar{R}(\beta)_{jk}^{jk}&=&-\frac{\displaystyle {\rm sh}
\left(\frac{n}{2\xi}\beta \right)}
{\displaystyle
{\rm sh}
\left(\frac{n}{2\xi}(\beta+\frac{2\pi i}{n}) \right)
},~~(j\neq k),\\
\bar{R}(\beta)_{jk}^{kj}&=&
\left\{
\begin{array}{cc}
\frac{\displaystyle
e^{-\frac{n}{2\xi}\beta}
{\rm sh}\left(\frac{\pi i}{\xi}\right)}
{
\displaystyle {\rm sh}
\left(\frac{n}{2\xi}(\beta+\frac{2\pi i}{n}) \right)
},& (j<k),\\
\frac{\displaystyle
e^{\frac{n}{2\xi}\beta}
{\rm sh}\left(\frac{\pi i}{\xi}\right)}
{
\displaystyle {\rm sh}
\left(\frac{n}{2\xi}(\beta+\frac{2\pi i}{n}) \right)
},& (j>k),
\end{array}
\right.
\end{eqnarray}
The $R$-matrix satisfies the Yang-Baxter equations,
\begin{eqnarray}
R_{12}^{VV}(\beta_1-\beta_2)
R_{13}^{VV}(\beta_1-\beta_3)
R_{23}^{VV}(\beta_2-\beta_3)=
R_{23}^{VV}(\beta_2-\beta_3)
R_{13}^{VV}(\beta_1-\beta_3)
R_{12}^{VV}(\beta_1-\beta_2),
\end{eqnarray}
and the unitarity,
\begin{eqnarray}
R_{12}^{VV}(\beta1-\beta_2)
R_{21}^{VV}(\beta_2-\beta_1)=id.
\end{eqnarray}
Let us set the monodromy matrix ${\cal T}(\beta)$
acting on the $(N+1)-$ fold tensor product
$V_0\otimes V_1 \otimes \cdots \otimes V_N, (V_k=V
={\mathbb{C}^n}
)$,
\begin{eqnarray}
{\cal T}(\beta)=R_{0,1}^{VV}(\beta)
\cdots R_{0,N}^{VV}(\beta)=
\left(
\begin{array}{ccccc}
A_{1,1} & A_{1,2} & \cdots & A_{1,n-1} & A_{1,n}\\
A_{2,1} & A_{2,2} & \cdots & A_{2,n-1} & A_{2,n}\\
\cdots & \cdots & \cdots & \cdots & \cdots \\
\cdots & \cdots & \cdots & \cdots & \cdots \\
A_{n-1,1} & A_{n-1,2} & \cdots & A_{n-1,n-1} & A_{n-1,n}\\
A_{n,1} & A_{n,2} & \cdots & A_{n,n-1} & A_{n,n}
\end{array}\right),
\end{eqnarray}
where the partition into $n \times n$ blocks is according to 
the base of $V_0$.
\\
Let us set the transfer matrix $T(\beta)$ 
acting on the $N$-th fold tensor product
$V_1\otimes \cdots \otimes V_N$
by
\begin{eqnarray}
T(\beta)=\sum_{j=1}^{n}A_{j,j}={\rm tr}_{V_0}({\cal T}(\beta)).
\end{eqnarray}
From the Yang-Baxter equation, we know the transfer matrices
commute each other.
\begin{eqnarray}
\left[T(\beta_1), T(\beta_2)\right]=0.
\end{eqnarray}
In the thermodynamic limit $N \to \infty$,
the logarithmic derivative of
the transfer matrix and the Hamiltonian (\ref{Hamiltonian})
have the following relation.
\begin{eqnarray}
\left(\frac{d}{d \beta}{\rm log}T\right)(0)
\sim {\cal H}.
\end{eqnarray}

~\\
A.Doikou and R.I.Nepomechie \cite{DN}
computed by Bethe Ansatz the scattering matrix
for the critical $A_{n-1}^{(1)}$ spin chain.
The scattering matrix $S^{}(\beta)$ is given by
\begin{eqnarray}
S^{}(\beta)=s(\beta)\bar{S}(\beta),~~s(\beta)=
\frac{S_2(-i\beta|\frac{2\pi}{n}(\xi-1),2\pi)
S_2(i\beta+\frac{2(n-1)\pi}{n}|\frac{2\pi}{n}(\xi-1),2\pi)
}{
S_2(i\beta|\frac{2\pi}{n}(\xi-1),2\pi)
S_2(-i\beta+\frac{2(n-1)\pi}{n}|\frac{2\pi}{n}(\xi-1),2\pi)
}.
\end{eqnarray}
The matrix $\bar{S}(\beta)$ is given as follows :
\begin{eqnarray}
\bar{S}(\beta)v_{k_1}\otimes v_{k_2}=
\sum_{j_1,j_2=0}^{n-1}
v_{j_1}\otimes v_{j_2}
\bar{S}(\beta)_{j_1 j_2}^{k_1 k_2},
\end{eqnarray}
where the nonzero entrise are 
\begin{eqnarray}
\bar{S}(\beta)_{jj}^{jj}&=&1,\\
\bar{S}(\beta)_{jk}^{jk}&=&-\frac{\displaystyle {\rm sh}
\left(\frac{n}{2(\xi-1)}\beta \right)}
{\displaystyle
{\rm sh}
\left(\frac{n}{2(\xi-1)}(\beta-\frac{2\pi i}{n}) \right)
},~~(j\neq k),\\
\bar{S}(\beta)_{jk}^{kj}&=&
\left\{
\begin{array}{cc}
-\frac{\displaystyle
e^{-\frac{n}{2(\xi-1)}\beta}
{\rm sh}\left(\frac{\pi i}{(\xi-1)}\right)}
{
\displaystyle {\rm sh}
\left(\frac{n}{2(\xi-1)}(\beta-\frac{2\pi i}{n}) \right)
},& (j<k),\\
-\frac{\displaystyle
e^{\frac{n}{2(\xi-1)}\beta}
{\rm sh}\left(\frac{\pi i}{(\xi-1)}\right)}
{
\displaystyle {\rm sh}
\left(\frac{n}{2(\xi-1)}(\beta-\frac{2\pi i}{n}) \right)
},& (j>k),
\end{array}
\right.
\end{eqnarray}
The scattering matrix of the present model
agrees with the scattering matrix
of the $A_{n-1}^{(1)}$ Toda fields theory
with imaginary coupling.

\section{Vertex Operators}

The type-I vertex operators
satisfy 
\begin{eqnarray}
\Phi_{j_2}(\beta_1)\Phi_{j_1}(\beta_2)
=\sum_{k_1 k_2=0}^{n-1}
R^{VV}(\beta_1-\beta_2)_{j_1 j_2}^{k_1 k_2}
\Phi_{k_1}(\beta_2)\Phi_{k_2}(\beta_1).\label{RVV}
\end{eqnarray}
The dual type-I vertex operators satisfy
\begin{eqnarray}
\Phi_{j_2}^*(\beta_1)\Phi_{j_1}^*(\beta_2)
=\sum_{k_1 k_2=0}^{n-1}
R^{V^*V^*}(\beta_1-\beta_2)_{j_1 j_2}^{k_1 k_2}
\Phi_{k_1}^*(\beta_2)\Phi_{k_2}^*(\beta_1),
\label{RV*V*}
\end{eqnarray}
where we set
\begin{eqnarray}
R^{V^*V^*}(\beta)_{j_1 j_2}^{k_1 k_2}=
R^{V V}(\beta)_{j_2 j_1}^{k_2 k_1}.
\end{eqnarray}
The type-I and dual type-I vertex operators
satisfy
\begin{eqnarray}
\Phi_{j_2}(\beta_1)\Phi_{j_1}^*\left(\beta_2\right)
=\sum_{k_1 k_2=0}^{n-1}
R^{V^* V}(\beta_1-\beta_2)_{j_1 j_2}^{k_1 k_2}
\Phi_{k_1}^*(\beta_2)\Phi_{k_2}(\beta_1),
\label{RV*V}
\end{eqnarray}
where we set
\begin{eqnarray}
R^{VV^*}(\beta)=r^*(\beta)\bar{R}^*(\beta),~~
r^*(\beta)=\frac
{
S_2(-i\alpha+\pi|\frac{2\pi}{n}\xi,2\pi)
S_2(i\alpha+\pi+\frac{2\pi}{n}|\frac{2\pi}{n}\xi,2\pi)
}
{S_2(i\alpha+\pi|\frac{2\pi}{n}\xi,2\pi)
S_2(-i\alpha+\pi+\frac{2\pi}{n}|\frac{2\pi}{n}\xi,2\pi)
}.
\end{eqnarray}
The matrix $\bar{R}^*(\beta)$ is given as follows :
\begin{eqnarray}
\bar{R}^*(\beta)v_{k_1}\otimes v_{k_2}=
\sum_{j_1,j_2=0}^{n-1}
v_{j_1}\otimes v_{j_2}
\bar{R}^*(\beta)_{j_1 j_2}^{k_1 k_2},
\end{eqnarray}
where the nonzero entrise are 
\begin{eqnarray}
\bar{R}^*(\beta)_{jk}^{jk}&=&1,~~~(j\neq k),\\
\bar{R}^*(\beta)_{jj}^{jj}&=&-\frac{\displaystyle {\rm sh}
\left(\frac{n}{2\xi}\beta \right)}
{\displaystyle
{\rm sh}
\left(\frac{n}{2\xi}(\beta+\frac{2\pi i}{n}) \right)
},\\
\bar{R}^*(\beta)_{jj}^{kk}&=&
\left\{
\begin{array}{cc}
\frac{\displaystyle
e^{-\frac{n}{2\xi}\beta}
{\rm sh}\left(\frac{\pi}{\xi}i\right)}
{
\displaystyle {\rm sh}
\left(\frac{n}{2\xi}(\beta+\frac{2\pi i}{n}) \right)
},& (j>k),\\
\frac{\displaystyle
e^{\frac{n}{2\xi}\beta}
{\rm sh}\left(\frac{\pi}{\xi}i\right)}
{
\displaystyle {\rm sh}
\left(\frac{n}{2\xi}(\beta+\frac{2\pi i}{n}) \right)
},& (j<k),
\end{array}
\right.
\end{eqnarray}
The type-I and the dual type-I vertex operators satisfy
\begin{eqnarray}
\Phi_{j_2}^*(\beta_1)\Phi_{j_1}(\beta_2)
=\sum_{k_1 k_2=0}^{n-1}
R^{V V^*}(\beta_1-\beta_2)_{j_1 j_2}^{k_1 k_2}
\Phi_{k_1}(\beta_2)\Phi_{k_2}^*(\beta_1),
\label{RVV*}
\end{eqnarray}
where we set
\begin{eqnarray}
R^{VV^*}(\beta)_{j_1 j_2}^{k_1 k_2}=
R^{V^*V}(\beta)_{k_1 k_2}^{j_1 j_2}.
\end{eqnarray}
The type-I vertex operator and it's dual
satisfy the inversion relation :
\begin{eqnarray}
\Phi_{j_1}(\beta)\Phi_{j_2}^*\left(\beta+\pi i
\right)=g_n^{-1}
e^{\frac{2\pi i}{n}j_1} 
\delta_{j_1,j_2} id,~~~(j_1\leq j_2).
\label{inversionI2}
\end{eqnarray}
where we set
\begin{eqnarray}
g_n^{-1}=
\left(\frac{2\xi}{n}\right)^{n-1}
{\rm sh}\left(\frac{\pi i}{\xi}\right)
e^{-\frac{\xi-1}{\xi}(\gamma+{\rm log}\frac{2\pi \xi}{n})n
+\frac{\pi i}{n}(1-n)}
\frac{\Gamma(1/\xi)^{n-1}}{\Gamma(1-1/\xi)}.
\end{eqnarray}

\section{Free boson realizations}

In this section we give the free boson realizations
of the vertex operators.\\
Let us set bose-fields as
\begin{eqnarray}
[b_j(t),b_k(t')]=-\frac{1}{t}
\frac{{\rm sh}\left(\frac{\pi}{n}(a_{j}|a_{k})t\right)
{\rm sh}\left(\frac{\pi}{n}(\xi-1) t\right)
}
{{\rm sh}\left(\frac{\pi}{n}t\right)
{\rm sh}
\left(\frac{\pi}{n}\xi t\right)
}
\delta(t+t')
\end{eqnarray}
Let us set
\begin{eqnarray}
b_1^*(t)&=&-\sum_{j=1}^{n-1}b_j(t)
\frac{{\rm sh}\frac{(n-j)\pi t}{n}}
{{\rm sh}\pi t},\\
b_{n-1}^*(t)&=&-\sum_{j=1}^{n-1}b_j(t)
\frac{{\rm sh}\frac{j\pi t}{n}}
{{\rm sh}\pi t}.
\end{eqnarray}
We have
\begin{eqnarray}
[b_1^*(t),b_j(t')]&=&
\delta_{j1}\frac{1}{t}\frac{{\rm sh}
\left(\frac{\pi}{n}(\xi-1) t\right)
}
{{\rm sh}\left(\frac{\pi}{n}\xi t\right)
}\delta(t+t'),\\
~[b_j(t),b_{n-1}^*(t')]&=&
\delta_{j,n-1}\frac{1}{t}\frac{{\rm sh}
\left(\frac{\pi}{n}(\xi-1) t\right)
}
{{\rm sh}\left(\frac{\pi}{n}\xi t\right)
}\delta(t+t').
\end{eqnarray}
Let us set the basic operators as
\begin{eqnarray}
U_j(\beta)&=&:\exp\left(
-\int_{-\infty}^\infty
b_j(t)e^{i\beta t}dt\right):~(1\leq j \leq n-1),\\
U_0(\beta)&=&
:\exp\left(-
\int_{-\infty}^\infty
b_1^*(t)e^{i\beta t}dt\right):,\\
U_n(\beta)&=&
:\exp\left(-
\int_{-\infty}^\infty
b_{n-1}^*(t)e^{i\beta t}dt\right):.
\end{eqnarray}
The bosonization of the type-I vertex operator is given by
\begin{eqnarray}
\Phi_j(\beta)&=&
\int_{-\infty}^\infty \frac{d\alpha_1}{2\pi i}
\cdots
\int_{-\infty}^\infty \frac{d\alpha_j}{2\pi i}
\frac{e^{\frac{n}{2\xi}(\alpha_j-\beta)}
U_0(\beta)U_1(\alpha_1)\cdots U_j(\alpha_j)}
{\prod_{k=1}^j
{\rm sh}\left(\frac{n}{2\xi}(\alpha_{k-1}-\alpha_{k}-
\frac{\pi i}{n})\right)}\\
&=&\int_{-\infty}^\infty \frac{d\alpha_1}{2\pi i}
\cdots
\int_{-\infty}^\infty \frac{d\alpha_j}{2\pi i}
\frac{e^{\frac{n}{2\xi}(\alpha_j-\beta)}
U_j(\alpha_j)U_{j-1}(\alpha_{j-1})\cdots U_0(\beta)}
{\prod_{k=1}^j
{\rm sh}\left(\frac{n}{2\xi}(\alpha_{k}-\alpha_{k-1}-
\frac{\pi i}{n})\right)},~(0\leq j \leq n-1)\nonumber
\end{eqnarray}
where $\alpha_0=\beta$.\\
The bosonization of the dual type-I vertex operator is given by
\begin{eqnarray}
\Phi_j^*(\beta)&=&
\int_{-\infty}^\infty \frac{d\alpha_{j+1}}{2\pi i}
\cdots
\int_{-\infty}^\infty \frac{d\alpha_{n-1}}{2\pi i}
\frac{e^{\frac{n}{2\xi}(\alpha_{j+1}-\beta)}
U_{j+1}(\alpha_{j+1})\cdots U_{n-1}(\alpha_{n-1})
U_n(\beta)}
{\prod_{k=j+1}^{n-1}
{\rm sh}\left(\frac{n}{2\xi}(\alpha_{k}-\alpha_{k+1}-
\frac{\pi i}{n})\right)},\\
&=&
\int_{-\infty}^\infty \frac{d\alpha_{j+1}}{2\pi i}
\cdots
\int_{-\infty}^\infty \frac{d\alpha_{n-1}}{2\pi i}
\frac{e^{\frac{n}{2\xi}(\alpha_{j+1}-\beta)}
U_n(\beta)U_{n-1}(\alpha_{n-1})\cdots
U_{j+1}(\alpha_{j+1})}
{\prod_{k=j+1}^{n-1}
{\rm sh}\left(\frac{n}{2\xi}(\alpha_{k+1}-\alpha_{k}-
\frac{\pi i}{n})\right)},~(0\leq j \leq n-1),\nonumber
\end{eqnarray}
where $\alpha_n=\beta$.
\\
T.Miwa
and Y.Takeyama
\cite{MT} gave the bosonization of some vertex operator,
to construct the solutions of the
critical quantum Knizhnik-Zamolodchikov equations
at Level $0$.
The calculation of $S$-matrix of
the critical $A_{n-1}^{(1)}$ chain
\cite{DN} teaches us
that their vertex operators are the dual type-II
vertex operators of the present model.

\section{Proof}
In this section we prove that
the bosonizations of the vertex operators satisfy
the commutation relations 
(\ref{RVV}), (\ref{RV*V*}), (\ref{RV*V}), (\ref{RVV*})
and the inversion relation (\ref{inversionI2}).
For convenience we list below formulae for the
contractions of the basic operators.

\begin{eqnarray}
U_j(\beta_1)
U_k(\beta_2)=U_k(\beta_2)
U_j(\beta_1),~~(|j-k|\geq 2).
\end{eqnarray}

\begin{eqnarray}
&&U_j(\beta_1)U_j(\beta_2)=
U_j(\beta_2)U_j(\beta_1)r(\beta_1-\beta_2),~~(j=0,n),\\
&&r(\beta)=
-\frac
{
S_2(i\alpha|\frac{2\pi}{n}\xi,2\pi)
S_2(-i\alpha+\frac{2\pi}{n}|\frac{2\pi}{n}\xi,2\pi)
}
{S_2(-i\alpha|\frac{2\pi}{n}\xi,2\pi)
S_2(i\alpha+\frac{2\pi}{n}|\frac{2\pi}{n}\xi,2\pi)
},
\end{eqnarray}

\begin{eqnarray}
U_0(\beta_1)U_{n-1}(\beta_2)=r^{*}(\beta_1-\beta_2)
U_{n-1}(\beta_2)U_{0}(\beta_1),\\
r^*(\beta)=\frac
{
S_2(-i\alpha+\pi|\frac{2\pi}{n}\xi,2\pi)
S_2(i\alpha+\pi+\frac{2\pi}{n}|\frac{2\pi}{n}\xi,2\pi)
}
{S_2(i\alpha+\pi|\frac{2\pi}{n}\xi,2\pi)
S_2(-i\alpha+\pi+\frac{2\pi}{n}|\frac{2\pi}{n}\xi,2\pi)
}.
\end{eqnarray}

\begin{eqnarray}
U_j(\beta_1)U_j(\beta_2)&=&H(\beta_1-\beta_2)
U_j(\beta_2)U_j(\beta_1),~~(1\leq j \leq n-1),\\
H(\beta)&=&-
\frac
{{\rm sh}\left(
\frac{n}{2\xi}(\beta+\frac{2\pi i}{n})\right)}
{{\rm sh}\left(
\frac{n}{2\xi}(-\beta+\frac{2\pi i}{n})\right)},
\end{eqnarray}

\begin{eqnarray}
U_j(\beta_1)U_{j-1}(\beta_2)&=&I(\beta_1-\beta_2)
U_{j-1}(\beta_2)U_j(\beta_1),~~(1\leq j \leq n),\\
I(\beta)&=&
\frac{{\rm sh}\left(\frac{n}{2\xi}\left(\beta-
\frac{\pi i}{n}\right)\right)}
{
{\rm sh}\left(\frac{n}{2\xi}\left(-\beta-
\frac{\pi i}{n}\right)\right)
}.
\end{eqnarray}
We are to prove the various properties of
the vertex operators
along the line of the papers \cite{AJMP, FKQ}.
We are to prove the commutation relation of the type-I
vertex operators (\ref{RVV}).
Consider the integral of the form :
\begin{eqnarray}
\int_{-\infty}^\infty 
\int_{-\infty}^\infty d\alpha_1 d\alpha_2
U_j(\alpha_1)U_j(\alpha_2)F(\alpha_1,\alpha_2).
\end{eqnarray}
Due to the commutation relation of $U_j(\alpha)$,
the above integral equals to
\begin{eqnarray}
\int_{-\infty}^\infty 
\int_{-\infty}^\infty d\alpha_1 d\alpha_2
U_j(\alpha_1)U_j(\alpha_2)F(\alpha_2,\alpha_1)
H(\alpha_2-\alpha_1)
\end{eqnarray}
Observing this we define 'weak equality' in the following sense.
We say that the fuctions $G_1(\alpha_1,\alpha_2)$
and $G_2(\alpha_1,\alpha_2)$ are equal in weak sense
if 
\begin{eqnarray}
G_1(\alpha_1,\alpha_2)+
H(\alpha_2-\alpha_1)G_1(\alpha_2,\alpha_1)=
G_2(\alpha_1,\alpha_2)+
H(\alpha_2-\alpha_1)G_2(\alpha_2,\alpha_1).
\end{eqnarray}
We write
\begin{eqnarray}
G_1(\alpha_1,\alpha_2)\sim
G_2(\alpha_1,\alpha_2).
\end{eqnarray}
To prove the commutation relations
(\ref{RVV}) and (\ref{RV*V*}) 
it is enough to prove the equalities of the integrand parts
in weakly sense.

Firsi we will show
\begin{eqnarray}
\Phi_\mu(\alpha_0)\Phi_\mu(\alpha_0')
=r(\alpha_0-\alpha_0')
\Phi_\mu(\alpha_0')\Phi_\mu(\alpha_0)~~~(0\leq \mu \leq n-1).
\label{RVVsame}
\end{eqnarray}
In what follows we use the notations :
\begin{eqnarray}
b(\alpha)=-\frac{{\rm sh}\rho \alpha}
{{\rm sh}\rho (\alpha+\frac{2\pi i}{n})},~~
c(\alpha)=\frac{
{\rm sh}\rho \frac{2\pi i}{n}}{
{\rm sh}\rho (\alpha+\frac{2\pi i}{n})
},~~d(\alpha)=
\frac{e^{\rho\alpha}}{{\rm sh}\rho(-\alpha-\frac{\pi i}{n})},
~~\rho=\frac{n}{2\xi}.
\end{eqnarray}
For $\mu=0$ case it is just the commutation relation
of $U_0(\alpha)$.
For $\mu\geq 1$
case we will show that by induction.
By using the commutation relations of the basic operators,
we can rearrange the operator part as
\begin{eqnarray}
U_0(\alpha_0)U_0(\alpha_0')
U_1(\alpha_1)U_1(\alpha_1')\cdots
U_{\mu}(\alpha_\mu)U_\mu(\alpha_\mu').
\end{eqnarray}
The integrand function of (LHS) of (\ref{RVVsame})
is given by
\begin{eqnarray}
L_\mu(\alpha_0 \alpha_0' 
\cdots \alpha_\mu \alpha_\mu')=
\prod_{k=1}^{\mu}d
(\alpha_k-\alpha_{k-1})
\prod_{k=1}^{\mu}d
(\alpha_k'-\alpha_{k-1}')
\prod_{k=1}^\mu I(\alpha_k-\alpha_{k-1}')
\end{eqnarray}
The integrand of (RHS) is given by
the exchange of variables $\alpha_0 \leftrightarrow \alpha_0'$
of (LHS).
\begin{eqnarray}
R_\mu(\alpha_0 \alpha_0' 
\cdots \alpha_\mu \alpha_\mu')=
L_\mu(\alpha_0' \alpha_0 
\cdots \alpha_\mu \alpha_\mu').
\end{eqnarray}
For $\mu=1$ case 
the weakely identity for variables
$\alpha_1, \alpha_1'$ can be shown by exact calculation.
\begin{eqnarray}
L_1(\alpha_0 \alpha_0' \alpha_1 \alpha_1')
\sim
R_1(\alpha_0 \alpha_0' \alpha_1 \alpha_1').
\end{eqnarray}
Because the integrand function 
splits into two parts :
\begin{eqnarray}
L_\mu(\alpha_0 \alpha_0'
\alpha_1 \alpha_1' \cdots \alpha_\mu \alpha_\mu')=
L_1(\alpha_0 \alpha_0' \alpha_1 \alpha_1')
L_{\mu-1}(\alpha_1 \alpha_1'
\alpha_2 \alpha_2' \cdots \alpha_\mu \alpha_\mu'),
\end{eqnarray}
the case $\mu \geq 2$ follows from induction.

Next we will show 
\begin{eqnarray}
&&\Phi_{\mu}(\alpha_0)\Phi_{\nu}(\alpha_0')\nonumber\\
&=&r(\alpha_0-\alpha_0')\left(b(\alpha_0-\alpha_0')
\Phi_{\nu}(\alpha_0')\Phi_{\mu}(\alpha_0)+
e^{{\rm sgn}(\nu-\mu)\rho(\alpha_0-\alpha_0')}
c(\alpha_0-\alpha_0')\Phi_{\mu}(\alpha_0')\Phi_{\nu}(\alpha_0)
\right).\label{RVVneq}
\end{eqnarray}
We prove the case $\nu>\mu$.
The case $\mu>\nu$ is similar.
For the case $\nu>\mu=0$,
the equality (\ref{RVVneq}) follows from the following
integrand
equality, which can be derived by direct calculations. 
\begin{eqnarray}
d(\alpha_1'-\alpha_0')=
b(\alpha_0-\alpha_0')I(\alpha_1'-\alpha_0)d(\alpha_1'-\alpha_0')
+c(\alpha_0-\alpha_0')e^{\rho(\alpha_0-\alpha_0')}
d(\alpha_1'-\alpha_0).
\end{eqnarray}
For the case $\nu>\mu\geq 1$,
the equality (\ref{RVVneq}) follows from the weak equality 
with respect to the variables $(\alpha_1 \alpha_1'),
(\alpha_2 \alpha_2')\cdots (\alpha_\mu \alpha_\mu')
$ :
\begin{eqnarray}
A_\mu(\alpha_0\alpha_0'\cdots \alpha_{\mu+1} \alpha_{\mu+1}') 
+B_\mu
(\alpha_0\alpha_0'\cdots \alpha_{\mu+1} \alpha_{\mu+1}') 
 +C_\mu
(\alpha_0\alpha_0'\cdots \alpha_{\mu+1} \alpha_{\mu+1}') \sim 0
,\label{eqn:ABC}
\end{eqnarray}
where we set
\begin{eqnarray}
A_\mu(\alpha_0 \alpha_0' \cdots \alpha_{\mu+1} \alpha_{\mu+1}')
=
\prod_{k=0}^{\mu-1} I(\alpha_{k+1}-\alpha_{k}')
\prod_{k=0}^{\mu-1} d(\alpha_{k+1}-\alpha_{k})
\prod_{k=0}^{\mu}
d(\alpha_{k+1}'-\alpha_{k}'),
\end{eqnarray}
\begin{eqnarray}
B_\mu(\alpha_0 \alpha_0' \cdots \alpha_{\mu+1} \alpha_{\mu+1}')
&=&-b(\alpha_0-\alpha_0')
\times
I(\alpha_{\mu+1}'-\alpha_\mu')
\left\{\prod_{k=1}^{\mu-1}
d(\alpha_{k+1}-\alpha_k')\right\}I(\alpha_1-\alpha_0)
\\
&\times&
d(\alpha_{\mu+1}'-\alpha_\mu)\left\{
\prod_{k=1}^{\mu-1}d(\alpha_{k+1}-\alpha_k)\right\}
d(\alpha_1-\alpha_0')
\left\{
\prod_{k=1}^{\mu-1}d(\alpha_{k+1}'-\alpha_k')\right\}
d(\alpha_1'-\alpha_0),\nonumber
\end{eqnarray}
and
\begin{eqnarray}
C_\mu(\alpha_0 \alpha_0' \cdots \alpha_{\mu+1} \alpha_{\mu+1}')
&=&-
c(\alpha_0-\alpha_0')e^{\rho(\alpha_0-\alpha_0')}
\times
\left\{\prod_{k=1}^{\mu-1}I(\alpha_{k+1}-\alpha_k')\right\}
I(\alpha_1-\alpha_0)\\
&\times&
\left\{\prod_{k=1}^\mu d(\alpha_{k+1}'-\alpha_k')\right\}
d(\alpha_1'-\alpha_0)
\left\{\prod_{k=1}^\mu d(\alpha_{k+1}-\alpha_k)\right\}
d(\alpha_1-\alpha_0').\nonumber
\end{eqnarray}
We are to prove the equation (\ref{eqn:ABC}) by induction of
$\mu$.
Inserting the equation (\ref{eqn:ABC}) for $\mu-1$ to
$B_\mu$, we have 
we have the equation in weakly sense with respect to
the variables $(\alpha_1 \alpha_1')
\cdots (\alpha_{\mu} \alpha_{\mu}')$ :
\begin{eqnarray}
B_\mu
(\alpha_0 \alpha_0' \cdots \alpha_{\mu+1} \alpha_{\mu+1}')
\sim
A'_\mu
(\alpha_0 \alpha_0' \cdots \alpha_{\mu+1} \alpha_{\mu+1}')
+
C'_\mu
(\alpha_0 \alpha_0' \cdots \alpha_{\mu+1} \alpha_{\mu+1}'),
\end{eqnarray}
where we set
\begin{eqnarray}
A'_\mu
(\alpha_0 \alpha_0' \cdots \alpha_\mu \alpha_{\mu+1}')=
-\frac{b(\alpha_0-\alpha_0')}{b(\alpha_1'-\alpha_1)}
\left\{
\prod_{k=2}^\mu d(\alpha_{k+1}'-\alpha_k')\right\}
d(\alpha_2'-\alpha_1)d(\alpha_1-\alpha_0')\nonumber\\
\times\left\{
\prod_{k=2}^{\mu-1} d(\alpha_{k+1}-\alpha_k)\right\}
d(\alpha_2-\alpha_1')d(\alpha_1'-\alpha_0)
\left\{
\prod_{k=2}^{\mu-1} I(\alpha_{k+1}-\alpha_k')\right\}
I(\alpha_2-\alpha_1)I(\alpha_1-\alpha_0),
\end{eqnarray}
and

\begin{eqnarray}
&&C'_\mu
(\alpha_0 \alpha_0' \cdots \alpha_{\mu+1} \alpha_{\mu+1}')=
\frac{b(\alpha_0-\alpha_0')c(\alpha_1'-\alpha_1)
e^{\rho(\alpha_1'-\alpha_1)}}{b(\alpha_1'-\alpha_1)}\\
&\times&\left\{\prod_{k=1}^\mu
d(\alpha_{k+1}'-\alpha_k')\right\}d(\alpha_1'-\alpha_0)
\left\{\prod_{k=1}^{\mu-1}
d(\alpha_{k+1}-\alpha_k)\right\}d(\alpha_1-\alpha_0')
\left\{\prod_{k=1}^{\mu-1}
I(\alpha_{k+1}-\alpha_k')\right\}I(\alpha_1-\alpha_0).
\nonumber
\end{eqnarray}
Using the following relation :
$
b(\alpha)H(\alpha)=b(-\alpha),
$
we have
\begin{eqnarray}
A_\mu'
(\alpha_0 \alpha_0' \cdots \alpha_{\mu+1} \alpha_{\mu+1}')
\sim
A_\mu''
(\alpha_0 \alpha_0' \cdots \alpha_{\mu+1} \alpha_{\mu+1}'),
\end{eqnarray}
where
\begin{eqnarray}
&&A_\mu''
(\alpha_0 \alpha_0' \cdots \alpha_{\mu+1} \alpha_{\mu+1}')\\
&=&
-\frac{b(\alpha_0-\alpha_0')}{b(\alpha_1'-\alpha_1)}
\left\{\prod_{k=0}^\mu d(\alpha_{k+1}'-\alpha_k')\right\}
\left\{\prod_{k=0}^\mu d(\alpha_{k+1}-\alpha_k)\right\}
\left\{\prod_{k=0}^{\mu-1} I(\alpha_{k+1}-\alpha_k')\right\}
I(\alpha_1'-\alpha_0).\nonumber
\end{eqnarray}
We have
\begin{eqnarray}
A_\mu+A_\mu''
&\sim&
\frac{
{\rm sh}
\left(\rho(\alpha_0'-\alpha_1'-\frac{\pi i}{n})\right)
{\rm sh}
\left(\rho(\alpha_0-\alpha_1+\frac{\pi i}{n})\right)
{\rm sh}
\left(\rho(\alpha_0+\alpha_1-\alpha_0'-\alpha_1')\right)
{\rm sh}
\left(\rho(\frac{2 \pi i}{n})\right)
}{
{\rm sh}\left(
\rho(\alpha_0'-\alpha_1-\frac{\pi i}{n})\right)
{\rm sh}
\left(\rho(\alpha_1'-\alpha_0+\frac{2\pi i}{n})\right)
{\rm sh}
\left(\rho(\alpha_0-\alpha_1'-\frac{\pi i}{n})\right)
{\rm sh}
\left(\rho(\alpha_0-\alpha_0'+\frac{2\pi i}{n})\right)
}\nonumber\\
&\times&\frac{1}{b(\alpha_1'-\alpha_1)}
\left\{\prod_{k=0}^\mu d(\alpha_{k+1}'-\alpha_k')\right\}
\left\{\prod_{k=0}^{\mu-1} d(\alpha_{k+1}-\alpha_k)\right\}
\left\{\prod_{k=1}^{\mu-1} I(\alpha_{k+1}-\alpha_k')\right\},
\end{eqnarray}
where we have used the relation :
\begin{eqnarray}
&&I(\alpha_1-\alpha_0')b(\alpha_1'-\alpha_1)-
I(\alpha_1'-\alpha_0)b(\alpha_0-\alpha_0')\\
&=&
\frac{
{\rm sh}
\left(\rho(\alpha_0'-\alpha_1'-\frac{\pi i}{n})\right)
{\rm sh}
\left(\rho(\alpha_0-\alpha_1+\frac{\pi i}{n})\right)
{\rm sh}
\left(\rho(\alpha_0+\alpha_1-\alpha_0'-\alpha_1')\right)
{\rm sh}
\left(\rho(\frac{2 \pi i}{n})\right)
}{
{\rm sh}\left(
\rho(\alpha_0'-\alpha_1-\frac{\pi i}{n})\right)
{\rm sh}
\left(\rho(\alpha_1'-\alpha_0+\frac{2\pi i}{n})\right)
{\rm sh}
\left(\rho(\alpha_0-\alpha_1'-\frac{\pi i}{n})\right)
{\rm sh}
\left(\rho(\alpha_0-\alpha_0'+\frac{2\pi i}{n})\right)
}.\nonumber
\end{eqnarray}
We have
\begin{eqnarray}
&&C_\mu+C_\mu'=-
b(\alpha_0-\alpha_0')
\frac{{\rm sh}\left(\rho(\frac{2\pi i}{n})\right)
{\rm sh}\left(\rho(\alpha_0+\alpha_1-\alpha_0'-\alpha_1'))\right)
}{
{\rm sh}\left(\rho(\alpha_1'-\alpha_1)\right)
{\rm sh}\left(\rho(\alpha_0-\alpha_0'))\right)
}
\\
&&\left\{\prod_{k=1}^\mu
d(\alpha_{k+1}'-\alpha_k')\right\}d(\alpha_1'-\alpha_0)
\left\{\prod_{k=1}^{\mu-1}
d(\alpha_{k+1}-\alpha_k)\right\}d(\alpha_1-\alpha_0')
\left\{\prod_{k=1}^{\mu-1}
I(\alpha_{k+1}-\alpha_k')\right\}I(\alpha_1-\alpha_0),\nonumber
\end{eqnarray}
where we have used the relation :
\begin{eqnarray}
\frac{c(\alpha_1'-\alpha_1)}{b(\alpha_1'-\alpha_1)}
e^{\rho(\alpha_1'-\alpha_1)}
-\frac{c(\alpha_0-\alpha_0')}{b(\alpha_0-\alpha_0')}
e^{\rho(\alpha_0-\alpha_0')}
=-\frac{{\rm sh}\left(\rho(\frac{2\pi i}{n})\right)
{\rm sh}\left(\rho(\alpha_0+\alpha_1-\alpha_0'-\alpha_1'))\right)
}{
{\rm sh}\left(\rho(\alpha_1'-\alpha_1)\right)
{\rm sh}\left(\rho(\alpha_0-\alpha_0'))\right)
}.
\end{eqnarray}
We arrive at
\begin{eqnarray}
A_\mu+B_\mu+C_\mu \sim A_{\mu}+A_{\mu}''
+C_{\mu}+C_{\mu}'\sim 0. 
\end{eqnarray}
Now we have proved the commutation relation
(\ref{RVV}).
As the same arguments as the above we can show the 
commutation relations (\ref{RV*V*}), (\ref{RV*V}) 
and (\ref{RVV*}).

Next we prove the inversion relations.\\
The bosonization of the type I vertex operator
is deformed as
\begin{eqnarray}
\Phi_j(\beta)&=&
\left(\frac{i}{\pi}\right)^j
e^{-j\frac{\xi-1}{\xi}(\gamma+{\rm log}\frac{2\pi \xi}{n})}
\int_{C_j}
\frac{d\alpha_j}{2\pi i}
\cdots
\int_{C_1}
\frac{d\alpha_1}{2\pi i}e^{\frac{n}{2\xi}(\alpha_j-\beta)}
\\
&\times&:U_0(\beta)\cdots U_j(\alpha_j):
\prod_{k=1}^j
\Gamma\left(\frac{n}{2\pi \xi}i(\alpha_k-\alpha_{k-1})
+\frac{1}{2\xi}\right)
\Gamma\left(-\frac{n}{2\pi \xi}i(\alpha_k-\alpha_{k-1})
+\frac{1}{2\xi}
\right),\nonumber
\end{eqnarray}
where $\alpha_0=\beta$.
Here the contour $C_k,~(k=1,\cdots,j)$
is taken $(-\infty,\infty)$ except that the poles
\begin{eqnarray}
\alpha_k-\alpha_{k-1}=\frac{\pi i}{n}+
\frac{2\pi \xi i}{n}l,~~(l \in \mathbb{N}\geq0),
\end{eqnarray}
of $
\Gamma\left(\frac{n}{2\pi \xi}i(\alpha_k-\alpha_{k-1})
+\frac{1}{2\xi}\right)
$ are above $C_k$ and the poles
\begin{eqnarray}
\alpha_k-\alpha_{k-1}=-\frac{\pi i}{n}-
\frac{2\pi \xi i}{n}l,~~(l \in \mathbb{N}\geq0),
\end{eqnarray}
of
$\Gamma\left(-\frac{n}{2\pi \xi}i(\alpha_k-\alpha_{k-1})
+\frac{1}{2\xi}
\right)$ are below $C_k$.
\\
The bosonization of the dual type-I vertex operator is 
deformed as
\begin{eqnarray}
\Phi_j^*(\beta)&=&
\left(\frac{i}{\pi}\right)^{n-j-1}
e^{-(n-j-1)\frac{\xi-1}{\xi}(\gamma+{\rm log}\frac{2\pi \xi}{n})}
\int_{C_{j+1}^*}
\frac{d\alpha_{j+1}}{2\pi i}
\cdots
\int_{C_{n-1}^*}
\frac{d\alpha_{n-1}}{2\pi i}e^{\frac{n}{2\xi}(\alpha_{j+1}-\beta)}
\\
&\times&:U_{j+1}(\alpha_{j+1})\cdots U_n(\beta):
\prod_{k=j+1}^{n-1}
\Gamma\left(\frac{n}{2\pi \xi}i(\alpha_{k+1}-\alpha_{k})
+\frac{1}{2\xi}\right)
\Gamma\left(-\frac{n}{2\pi \xi}i(\alpha_{k+1}-\alpha_{k})
+\frac{1}{2\xi}
\right),\nonumber
\end{eqnarray}
where $\alpha_n=\beta$.
Here the contour $C_k^*,~(k=j+1,\cdots,n-1)$
is taken $(-\infty,\infty)$ except that the poles
\begin{eqnarray}
\alpha_k-\alpha_{k+1}=\frac{\pi i}{n}+
\frac{2\pi \xi i}{n}l,~~(l \in \mathbb{N}\geq0),
\end{eqnarray}
of $
\Gamma\left(-\frac{n}{2\pi \xi}i(\alpha_{k+1}-\alpha_{k})
+\frac{1}{2\xi}\right)
$ are above $C_k^*$ and the poles
\begin{eqnarray}
\alpha_k-\alpha_{k+1}=-\frac{\pi i}{n}-
\frac{2\pi \xi i}{n}l,~~(l \in \mathbb{N}\geq0),
\end{eqnarray}
of
$\Gamma\left(\frac{n}{2\pi \xi}i(\alpha_{k+1}-\alpha_{k})
+\frac{1}{2\xi}
\right)$ are below $C_k^*$.

Let us prove the relation (\ref{inversionI2}).
For $j_1<j_2$ we have
\begin{eqnarray}
\Phi_{j_1}(\beta_1)\Phi_{j_2}^*(\beta_2)
=r^*(\beta_1-\beta_2)
\Phi_{j_2}^*(\beta_2)\Phi_{j_1}(\beta_1).
\end{eqnarray}
As $\beta_2 \to \beta_1+\pi i$,
$r^*(\beta_1-\beta_2)$ becomes zero.\\
Next we prove the case $j_1=j_2$.
We have
\begin{eqnarray}
\Phi_j(\beta_1)\Phi_j^*(\beta_2)&=&
r^*(\beta_1-\beta_2)\int_{C_1}\frac{d\alpha_1}{2\pi i}
\cdots \int_{C_j}\frac{d\alpha_j}{2\pi i}
\int_{C_{j+1}^*}\frac{d\alpha_{j+1}}{2\pi i}
\cdots \int_{C_{n-1}^*}\frac{d\alpha_{n-1}}{2\pi i}
\nonumber\\
&\times&
U_n(\beta_2)U_{n-1}(\alpha_{n-1})
\cdots U_1(\alpha_1)U_0(\beta_1)\\
&\times&
e^{\frac{n}{2\xi}(\alpha_j+\alpha_{j+1}-
\beta_1-\beta_2)}
{\rm sh}\left(\frac{n}{2\xi}
(\alpha_{j}-\alpha_{j+1}-\frac{\pi i}{n})\right)
\prod_{k=1}^{n}\frac{1}{{\rm sh}\left(\frac{n}{2\xi}
(\alpha_k-\alpha_{k-1}-\frac{\pi i}{n})\right)}.
\nonumber
\end{eqnarray}
When we take the limit $\beta_2 \to \beta_1+\pi i$,
the contour is pinched but the function $r^*(\beta_1-\beta_2)$
has a zero.
Therefore the limit is evaluated by successively taking
the residues at $\alpha_k=\alpha_{k-1}+\frac{\pi i}{n}$
for $k=1,\cdots, j$,
and $\alpha_{k}=\alpha_{k+1}-\frac{\pi i}{n}$ 
for $k=j+1,\cdots, n-1$, successively.
The following relation is usefull.
\begin{eqnarray}
U_n(\beta+\pi i)U_{n-1}\left(\beta+\frac{n-1}{n}\pi i\right)
\cdots
U_1\left(\beta+\frac{\pi i}{n}\right)U_0(\beta)
=e^{-\frac{\xi-1}{\xi}(\gamma+{\rm log}\frac{2\pi \xi}{n})n}
\Gamma(1/\xi)^n.
\end{eqnarray}
Now we have proved the inversion relation (\ref{inversionI2}).

\section{Correlation functions}

In this section we derive a solution of the system of
difference equations (\ref{qKZ1}) and (\ref{qKZ2}),
algebraically, and obtain an integral
representation of it.\\
Let us introduce the degree operator $D$ on the Fock space :
\begin{eqnarray}
D~b_j(-t)|0\rangle=t~b_j(-t)|0\rangle,~(t>0).
\end{eqnarray}
We have
\begin{eqnarray}
e^{\lambda D}b(t)e^{-\lambda D}=
e^{-\lambda t}b(t).
\end{eqnarray}
Therefore the degree operator $D$ has the homogeneity
condition.
\begin{eqnarray}
e^{\lambda D}U_j(\beta)e^{-\lambda D}=U_j(\beta+i\lambda).
\end{eqnarray}
The vertex operator and the degree operator enjoy
the homogeneity property.
\begin{eqnarray}
e^{-\lambda D}\Phi_j(\beta)e^{\lambda D}=\
\Phi_j(\beta+i\lambda),~~
e^{-\lambda D}\Phi_j^*
(\beta)e^{\lambda D}=\Phi_j^*(\beta+i\lambda).
\label{Hom}
\end{eqnarray}
Now let us consider the trace functions for $\lambda>0$
defined by
\begin{eqnarray}
&&G(\beta_1 \cdots \beta_N |
\beta_{N+1} \cdots \beta_{2N}
)_{\epsilon_1 \cdots \epsilon_{2N}}
\nonumber\\
&=&
\frac{{\rm tr}_{\cal H}\left(e^{-\lambda D}
\Phi_{\epsilon_1}^*(\beta_1)
\cdots \Phi_{\epsilon_N}^*(\beta_N)
\Phi_{\epsilon_{N+1}}(\beta_{N+1})
\cdots \Phi_{\epsilon_{2N}}(\beta_{2N})
\right)}
{{\rm tr}_{\cal H}\left(e^{-\lambda D}\right)
},\label{tr}
\end{eqnarray}
where the space ${\cal H}$ is the Fock space of
the free bosons.

By using the homogeneity
condition (\ref{Hom}) and the commutation relations
(\ref{RVV}), (\ref{RV*V*}), (\ref{RVV*}),
it is shown that the above trace function satisfies
the desired difference equations (\ref{qKZ1}), (\ref{qKZ2}).
The inversion relation (\ref{inversionI2})
means the restriction equation
 (\ref{restriction}).

Using the free boson realizations,
we shall treat the trace of the form :
${\rm tr}_{\cal H}(e^{-\lambda D}{\cal O})$.
The calculation is simplified by the technique of Clavelli and 
Shapiro \cite{CS}.
Their prescription is as follows.
Introduce a copy of the bosons $a(t)~(t \in {\mathbb{R}})$
satisfying the relation $[a(t),b(t')]=0$ and the same commutation
relation as the $b(t)$.
Let 
\begin{eqnarray}
\tilde{b}(t)=\frac{b(t)}{1-e^{-\lambda t}}+a(-t),~~
\tilde{b}(-t)=\frac{a(t)}{e^{\lambda t}-1}+b(-t),~~(t>0).
\end{eqnarray}
For a linear operator ${\cal O}$ on the Fock space 
${\cal H}_b={\cal F}[b(-t)]$,
let $\tilde{\cal O}$ be the operator on 
${\cal H}_b \otimes {\cal H}_a, ~({\cal H}_a={\cal F}[a(-t)]
)$ obtaind by substituting $\tilde{b}(t)$
for $b(t)$.
We have then
\begin{eqnarray}
\frac{
\displaystyle 
{\rm tr}_{{\cal H}_b}\left(e^{-\lambda D} 
{\cal O}
\right)
}
{\displaystyle
{\rm tr}_{{\cal H}_b}(e^{-\lambda D})
}=
\langle \tilde{0} | \tilde{\cal O} | \tilde{0} \rangle,
\end{eqnarray}
where the left hand side
denotes the usual expectation value
with respect to the Fock vacuum
$| \tilde{0} \rangle= |0\rangle \otimes |0\rangle$,
$ \langle \tilde{0} | \tilde{0} \rangle$.
In what follows we use the following abbreviation :
\begin{eqnarray}
\langle {\cal O} \rangle_\lambda
=\frac{{\rm tr}_{{\cal H}}\left(e^{-\lambda D} 
{\cal O}
\right)
}
{\displaystyle
{\rm tr}_{\cal H}(e^{-\lambda D})
}.
\end{eqnarray}
We have the following.
\begin{eqnarray}
\langle b_j(t)b_k(t') \rangle_\lambda=
\frac{e^{\lambda t}}{e^{\lambda t}-1}
[b_j(t),b_k(t')].
\end{eqnarray}
We have the following formula usefull to
evaluate the function (\ref{tr}).
\begin{eqnarray}
\frac{
\displaystyle 
{\rm tr}_{H}\left(e^{-\lambda D} 
:e^{\int_{-\infty}^\infty
c(t)e^{i\beta_1t}dt}:
:e^{
\int_{-\infty}^\infty
d(t)e^{i\beta_1t}dt}:
\right)
}
{\displaystyle
{\rm tr}_{H}(e^{-\lambda D})
}=
\exp\left(
\int_0^\infty
A(t)\frac{{\rm ch}(i(\beta_1-\beta_2)+\frac{\lambda}{2})t}
{{\rm sh}\frac{\lambda t}{2}}dt\right),
\end{eqnarray}
where $c(t)$ and $d(t)$ are bosons satisfying
$[c(t),d(t')]=A(t)\delta(t+t')$ and $A(t)=-A(-t)$.
In order to understand an integral of the right hand side,
see the Appendix B.

The basic trace functions are evaluated as following.
\begin{eqnarray}
\langle U_0(\alpha_1)U_0(\alpha_2) \rangle_\lambda
=
\langle U_n(\alpha_1)U_n(\alpha_2) \rangle_\lambda
={\rm Const.}E_\lambda(\alpha_1-\alpha_2),
\end{eqnarray}
\begin{eqnarray}
\langle U_0(\alpha_1)U_n(\alpha_2) \rangle_\lambda
=\langle U_n(\alpha_1)U_0(\alpha_2) \rangle_\lambda
={\rm Const.}
E_\lambda^*(\alpha_1-\alpha_2).
\end{eqnarray}
Here we set
\begin{eqnarray}
E_\lambda(\alpha)=
\frac{S_3(-i\alpha+\frac{2\pi}{n})
S_3(i\alpha+\frac{2\pi}{n}+\lambda)}{
S_3(-i\alpha+\frac{2\pi}{n}\xi)
S_3(i\alpha+\frac{2\pi}{n}\xi+\lambda)
}\\
E_\lambda^*(\alpha)=
\frac{S_3(-i\alpha+\pi)S_3(i\alpha+\lambda+\pi)}{
S_3(-i\alpha+\frac{2\pi}{n}+\pi)
S_3(i\alpha+\lambda+\pi+\frac{2\pi}{n})},
\end{eqnarray}
where
\begin{eqnarray}
S_3(\beta)=S_3\left(\beta
\left|2\pi, \frac{2\pi}{n}\xi, \lambda
\right.\right).
\end{eqnarray}
\begin{eqnarray}
&&\langle U_j(\alpha_1)U_{j-1}(\alpha_2) \rangle_\lambda
=\langle U_{j-1}(\alpha_1)U_{j}(\alpha_2) \rangle_\lambda
\nonumber
\\
&=&{\rm Const.}
\varphi
(\alpha_1-\alpha_2)\times
\frac{1}
{
\displaystyle
{\rm sh}\left(\frac{n}{2\xi}(\alpha_1-\alpha_2
-\frac{\pi i}{n}+\lambda i)\right)},
\end{eqnarray}
\begin{eqnarray}
&&\langle U_{j}(\alpha_1)U_j(\alpha_2) \rangle_\lambda
={\rm Const.}\psi(\alpha_1-\alpha_2)\times\\
&\times&{\rm sh}\left(\frac{n}{2\xi}\alpha
\right)
{\rm sh}\left(\frac{n}{2\xi}(\alpha+\frac{2\pi i}{n})\right)
{\rm sh}\left(\frac{\pi}{\lambda}(\alpha+\frac{2\pi i}{n})\right)
{\rm sh}\left(\frac{\pi}{\lambda}
(-\alpha+\frac{2\pi i}{n})\right).\nonumber
\end{eqnarray}
Here we set
\begin{eqnarray}
\varphi(\alpha)&=&\frac{1}{\displaystyle
S_2\left(
\left.i\alpha+\frac{\pi}{n}\right|
\lambda, \frac{2\pi}{n}\xi\right)
S_2\left(\left.-i\alpha+\frac{\pi}{n}
\right|\lambda, \frac{2\pi}{n}\xi
\right)},
\\
\psi(\alpha)&=&
\frac{1}{
\displaystyle S_2\left(\left.
i\alpha-\frac{2\pi}{n}\right|
\lambda, \frac{2\pi}{n}\xi
\right)
S_2\left(\left.
-i\alpha-\frac{2\pi}{n}\right|
\lambda, \frac{2\pi}{n}\xi
\right)}.
\end{eqnarray}
The functions $\varphi(\alpha)$ and $\psi(\alpha)$
become the integral kernel of the correlation functions.\\
The function $\varphi(\alpha)$
has poles at
\begin{eqnarray}
\alpha=\pm i\left(n_1 \lambda +n_2 \frac{2\pi}{n}\xi
+\frac{\pi}{n}
\right),~~(n_1,n_2 \geq 0).
\end{eqnarray}
The function $\psi(\alpha)$
has poles at 
\begin{eqnarray}
\alpha=\pm i\left(n_1 \lambda +n_2 \frac{2\pi}{n}\xi
-\frac{2\pi}{n}
\right),~~(n_1,n_2 \geq 0).
\end{eqnarray}

The trace of the vertex operators (\ref{tr}) 
is evaluated by applying
the Wick's theorem.\\
The one-point correlation functions 
are evaluated as follows.
In what follows we set $\rho=\frac{n}{2\xi}$.
\begin{eqnarray}
G(\beta_1 |\beta_1')
_{\epsilon,\epsilon}
&=&
E_\lambda^*(\beta_1-\beta_1')
\int_{-\infty}^\infty \frac{d\alpha_1}{2\pi i}\cdots
\int_{-\infty}^\infty \frac{d\alpha_{n-1}}{2\pi i}
\prod_{k=1}^n\varphi(\alpha_k-\alpha_{k-1})
e^{\rho(\alpha_\epsilon+
\alpha_{\epsilon+1}-\beta_1-\beta_1')}\\
&\times&
{\rm sh}\left(\rho (\alpha_{\epsilon+1}-
\alpha_\epsilon-\frac{\pi i}{n})\right)
\prod_{k=1}^n
\frac{1}{\displaystyle
{\rm sh}\left(\rho(\alpha_k-\alpha_{k-1}-\frac{\pi i}{n})\right)
{\rm sh}\left(\rho(\alpha_k-\alpha_{k-1}-\frac{\pi i}{n}
+\lambda i
)\right)},\nonumber
\end{eqnarray}
where $\alpha_n=\beta_1,~\alpha_0=\beta_1'$.
Here we omit an irrelevant constant factor.
\\
The $N$ point correlation functions are evaluated
as follows.
\begin{eqnarray}
G(\beta_1\cdots \beta_N|\beta_N' \cdots \beta_1')
_{\epsilon_1 \cdots \epsilon_N,\epsilon_N,\cdots,
\epsilon_1}=
E(\beta_1 \cdots \beta_N|\beta_N' \cdots \beta_1')
\prod_{j,r} \int_{-\infty}^\infty
\frac{d\alpha_{j,r}}{2\pi i}I_{\epsilon_1 \cdots \epsilon_N}
(\left\{\alpha_{j,r}\right\}). 
\end{eqnarray}
We associate the variables $\alpha_{j,r},~(1\leq r \leq N,
0\leq j \leq \epsilon_r)$ to the basic operator 
$U_j(\alpha_{j,r})$
contained in the vertex operators 
$\Phi_{\epsilon_r}(\beta_r')$ and
the variables $\alpha_{j,r},~(1\leq r \leq N,
\epsilon_r+1\leq j \leq n)$ to the basic operator 
$U_j(\alpha_{j,r})$
contained in the vertex operators  
$\Phi_{\epsilon_r}^*(\beta_r)$.
\\
We set
\begin{eqnarray}
\alpha_{n,r}=\beta_r,&& \alpha_{0,r}=\beta_r',\\
{\cal N}_j^*=\{k|\epsilon_k \leq j-1 \},&&
{\cal N}_j=\{k|\epsilon_k \geq j \}.
\end{eqnarray}
The function $E(\beta_1\cdots \beta_N|\beta_N'
\cdots \beta_1')$ is given by
\begin{eqnarray}
&&E(\beta_1\cdots \beta_N|\beta_N'
\cdots \beta_1')\nonumber
\\
&=&
e^{-\rho(\beta_1+\cdots+\beta_N+
\beta_1'+\cdots+\beta_N')}
\prod_{1\leq j<k \leq N}E_\lambda(\beta_j-\beta_k)
\prod_{1\leq j<k \leq N}E_\lambda(\beta_k'-\beta_j')
\prod_{j,k=1}^N E_\lambda^*(\beta_j-\beta_k').
\end{eqnarray}
The integrand is given by
\begin{eqnarray}
I_{\epsilon_1 \cdots \epsilon_N}(
\{\alpha_{j,r}\})
=K(\{\alpha_{j,r}\})g(\{\alpha_{j,r}\})
h(\{\alpha_{j,r}\}).
\end{eqnarray}
Here the integral kernel is given by
\begin{eqnarray}
K(\{\alpha_{j,r}\})=
\prod_{j=1}^{n-1}
\prod_{r,s=1 \atop{r>s} }^N
\psi(\alpha_{j,r}-\alpha_{j,s})
\times
\prod_{j=1}^{n-1} \prod_{k,l=1}^{N}
\left\{\prod_{k \in {N_j^*} \atop{l \in {N_{j-1}}}}
\varphi(
\alpha_{j,k}-\alpha_{j,l})
\prod_{k \in {N_{j-1}^*} \atop{l \in N_{j}}}\varphi(
\alpha_{j,k}-\alpha_{j,l})
\right\}\nonumber\\
\times
\prod_{j=1}^{n-1} \prod_{k,l=1}^{N}
\left\{\prod_{k \in {N_j} \atop{l \in {N_{j-1}}}}
\varphi(
\alpha_{j,k}-\alpha_{j,l})
\prod_{k \in {N_{j}^*} \atop{l \in N_{j-1}^*}}\varphi(
\alpha_{j,k}-\alpha_{j,l})
\right\}^2.
\end{eqnarray}
Here we set
\begin{eqnarray}
h(\{\alpha_{j,r}\})
=\prod_{j=1}^{n-1}
\prod_{r,s=1 \atop{r>s}}^{N}
\left\{
{\rm sh}\left(\frac{\pi}{\lambda}
(\alpha_{j,r}-\alpha_{j,s}+
\frac{2\pi i}{n})\right)
{\rm sh}\left(\frac{\pi}{\lambda}
(\alpha_{j,s}-\alpha_{j,r}+
\frac{2\pi i}{n})\right)\right\},
\end{eqnarray}
and
\begin{eqnarray}
&&g(\{\alpha_{j,r}\})\nonumber\\
&=&
\prod_{j=1}^{n-1}
\prod_{r>s \atop{s \in N_j}}
\left\{
{\rm sh}\left(\rho
(\alpha_{j,r}-\alpha_{j,s})\right)
{\rm sh}\left(\rho
(\alpha_{j,r}-\alpha_{j,s}
+\frac{2\pi i}{n})\right)
\right\}\nonumber\\
&\times&
\prod_{j=1}^{n-1}
\prod_{r>s \atop{s \in N_j^*}}
\left\{
{\rm sh}\left(\rho
(\alpha_{j,s}-\alpha_{j,r})\right)
{\rm sh}\left(\rho
(\alpha_{j,s}-\alpha_{j,r}
+\frac{2\pi i}{n})\right)
\right\}\nonumber\\
&\times&
\prod_{j=1}^{n}
\prod_{r \in N_j \atop{s \in N_{j-1}}}
\left\{
{\displaystyle
{\rm sh}\left(\rho(\alpha_{j,r}-\alpha_{j-1,s}
-\frac{\pi i}{n}+\lambda i)\right)
{\rm sh}\left(\rho(\alpha_{j-1,r}-\alpha_{j,s}
-\frac{\pi i}{n}+\lambda i)\right)}
\right\}^{-1}\nonumber\\
&\times&
\prod_{j=1}^{n}
\prod_{r \in N_j^* \atop{s \in N_{j-1}^*}}
\left\{
{\displaystyle
{\rm sh}\left(\rho(\alpha_{j,r}-\alpha_{j-1,s}
-\frac{\pi i}{n}+\lambda i)\right)
{\rm sh}\left(\rho(\alpha_{j-1,r}-\alpha_{j,s}
-\frac{\pi i}{n}+\lambda i)\right)}
\right\}^{-1}\\
&\times&
\prod_{j=1}^{n}\left\{
\prod_{r \in N_j^* \atop{s \in N_{j-1}}}
{\displaystyle
{\rm sh}\left(\rho(\alpha_{j,r}-\alpha_{j-1,s}
-\frac{\pi i}{n}+\lambda i)\right)
\prod_{r\in N_{j-1}^*
\atop{s \in N_j}}
{\rm sh}\left(\rho(\alpha_{j,r}-\alpha_{j-1,s}
-\frac{\pi i}{n}+\lambda i)\right)}
\right\}^{-1}\nonumber\\
&\times&
\prod_{r=1}^N
e^{\rho(\alpha_{\epsilon_r}+
\alpha_{\epsilon_r+1})}
\prod_{r=1}^N
{\rm sh}\left(\rho(\alpha_{\epsilon_r+1,r}
-\alpha_{\epsilon_r,r}-\frac{\pi i}{n})\right)
\prod_{j=1}^{n}\prod_{r=1}^N
\left\{
{\rm sh}\left(\rho(\alpha_{j,r}-
\alpha_{j-1,r}-\frac{\pi i}{n})\right)
\right\}^{-1}.\nonumber
\end{eqnarray}
Here we omit an irrelevant constant factor.

~\\
~\\
{\bf Acknowledgements.}
~~
This work was partly supported by 
Grant-in-Aid for Encouragements for Young Scientists (A)
from Japan Society for the Promotion of Science. (11740099)

\begin{appendix}
\section{Multi-Gamma functions}

Here we summarize the multiple gamma and the multiple sine
functions, following Kurokawa \cite{K}.\\
Let us set the functions
$\Gamma_1(x|\omega), \Gamma_2(x|\omega_1, \omega_2)
$ and
$\Gamma_3(x|\omega_1, \omega_2, \omega_3)
$
by
\begin{eqnarray}
{\rm log}\Gamma_1(x|\omega)+\gamma B_{11}(x|\omega)&=&
\int_C\frac{dt}{2\pi i t}e^{-xt}
\frac{{\rm log}(-t)}{1-e^{-\omega t}},\\
{\rm log}\Gamma_2(x|\omega_1, \omega_2)
-\frac{\gamma}{2} B_{22}(x|\omega_1, \omega_2)&=&
\int_C\frac{dt}{2\pi i t}e^{-xt}
\frac{{\rm log}(-t)}{(1-e^{-\omega_1 t})
(1-e^{-\omega_2 t})},\\
{\rm log}\Gamma_3(x|\omega_1, \omega_2, \omega_3)
+\frac{\gamma}{3!} B_{33}(x|\omega_1, \omega_2, \omega_3)&=&
\int_C\frac{dt}{2\pi i t}e^{-xt}
\frac{{\rm log}(-t)}{(1-e^{-\omega_1 t})
(1-e^{-\omega_2 t})
(1-e^{-\omega_3 t})},
\end{eqnarray}
where
the functions $B_{jj}(x)$ are the multiple Bernoulli polynomials
defined by
\begin{eqnarray}
\frac{t^r e^{xt}}{
\prod_{j=1}^r (e^{\omega_j t}-1)}=
\sum_{n=0}^\infty
\frac{t^n}{n!}B_{r,n}(x|\omega_1 \cdots \omega_r),
\end{eqnarray}
more explicitly
\begin{eqnarray}
B_{11}(x|\omega)&=&\frac{x}{\omega}-\frac{1}{2},\\
B_{22}(x|\omega)&=&\frac{x^2}{\omega_1 \omega_2}
-\left(\frac{1}{\omega_1}+\frac{1}{\omega_2}\right)x
+\frac{1}{2}+\frac{1}{6}\left(\frac{\omega_1}{\omega_2}
+\frac{\omega_2}{\omega_1}\right).
\end{eqnarray}
Here $\gamma$ is Euler's constant,
$\gamma=\lim_{n\to \infty}
(1+\frac{1}{2}+\frac{1}{3}+\cdots+\frac{1}{n}-{\rm log}n)$.\\
Here the contor of integral is given by

~\\
~\\

\unitlength 0.1in
\begin{picture}(34.10,11.35)(17.90,-19.35)
%
\special{pn 8}%
\special{pa 5200 800}%
\special{pa 2190 800}%
\special{fp}%
\special{sh 1}%
\special{pa 2190 800}%
\special{pa 2257 820}%
\special{pa 2243 800}%
\special{pa 2257 780}%
\special{pa 2190 800}%
\special{fp}%
\special{pa 2190 1600}%
\special{pa 5190 1600}%
\special{fp}%
\special{sh 1}%
\special{pa 5190 1600}%
\special{pa 5123 1580}%
\special{pa 5137 1600}%
\special{pa 5123 1620}%
\special{pa 5190 1600}%
\special{fp}%
%
\special{pn 8}%
\special{pa 5190 1200}%
\special{pa 2590 1210}%
\special{fp}%
\put(25.9000,-12.1000){\makebox(0,0)[lb]{$0$}}%
%
\special{pn 8}%
\special{ar 2190 1210 400 400  1.5707963 4.7123890}%
\put(33.9000,-20.2000){\makebox(0,0){{\bf Contour} $C$}}%
\end{picture}%

~\\
~\\

Let us set
\begin{eqnarray}
S_1(x|\omega)&=&\frac{1}{\Gamma_1(\omega-x|\omega)
\Gamma_1(x|\omega)},\\
S_2(x|\omega_1,\omega_2)&=&\frac{
\Gamma_2(\omega_1+\omega_2-x|\omega_1,\omega_2)}{
\Gamma_2(x|\omega_1,\omega_2)},\\
S_3(x|\omega_1,\omega_2,\omega_3)&=&\frac{1}{
\Gamma_3(\omega_1+\omega_2+\omega_3-x|\omega_1,\omega_2,\omega_3)
\Gamma_3(x|\omega_1,\omega_2,\omega_3)}
\end{eqnarray}
We have
\begin{eqnarray}
\Gamma_1(x|\omega)=e^{(\frac{x}{\omega}-\frac{1}{2}){\rm log}
\omega}\frac{\Gamma(x/\omega)}{\sqrt{2\pi}},~
S_1(x|\omega)=2{\rm sin}(\pi x/\omega),
\end{eqnarray}
\begin{eqnarray}
\frac{\Gamma_2(x+\omega_1|\omega_1,\omega_2)}{
\Gamma_2(x|\omega_1,\omega_2)}=\frac{1}{\Gamma_1(x|\omega_2)},~
\frac{S_2(x+\omega_1|\omega_1,\omega_2)}{
S_2(x|\omega_1,\omega_2)}=\frac{1}{S_1(x|\omega_2)},~
\frac{\Gamma_1(x+\omega|\omega)}{\Gamma_1(x|\omega)}=x.
\end{eqnarray}

\begin{eqnarray}
\frac{
\Gamma_3(x+\omega_1|\omega_1,\omega_2, \omega_3
}{\Gamma_3(x|\omega_1,\omega_2, \omega_3)}
=\frac{1}{\Gamma_2(x|\omega_2, \omega_3)},~
\frac{S_3(x+\omega_1|\omega_1,\omega_2,\omega_3)}{
S_3(x|\omega_1,\omega_2,\omega_3)}=\frac{1}{S_2(x|\omega_2,
\omega_3)}.
\end{eqnarray}

\begin{eqnarray}
{\rm log}S_2(x|\omega_1 \omega_2)
=\int_C \frac{{\rm sh}(x-\frac{\omega_1+\omega_2}{2})t}
{2{\rm sh}\frac{\omega_1 t}{2}
{\rm sh}\frac{\omega_2 t}{2}
}{\rm log}(-t)\frac{dt}{2\pi i t},~(0<{\rm Re}x<
\omega_1+\omega_2).
\end{eqnarray}

\begin{eqnarray}
S_2(x|\omega_1 \omega_2)=
\frac{2\pi}{\sqrt{\omega_1 \omega_2}}x +O(x^2),~~(x \to 0).
\end{eqnarray}

\begin{eqnarray}
S_2(x|\omega_1 \omega_2)
S_2(-x|\omega_1 \omega_2)=-4
{\rm sin}\frac{\pi x}{\omega_1}
{\rm sin}\frac{\pi x}{\omega_2}.
\end{eqnarray}

\section{Normal Ordering}

Here we list the formulas of the form
\begin{eqnarray}
X(\beta_1)Y(\beta_2)=C_{XY}(\beta_1-\beta_2)
:X(\beta_1)X(\beta_2):,
\end{eqnarray}
where $X,Y=U_j$, and $C_{XY}(\beta)$ is a meromorphic
function on ${\mathbb{C}}$.
These formulae follow from the commutation relation
of the free bosons.
When we compute the contraction of the basic
operators,
we often encounter an integral
\begin{eqnarray}
\int_0^\infty
F(t)dt,
\end{eqnarray}
which is divergent at $t=0$.
Here we adopt the following prescription
for regularization :
it should be understood as the countour integral,
\begin{eqnarray}
\int_C F(t)\frac{{\rm log}(-t)}{2\pi i}dt, 
\end{eqnarray}
where the countour $C$ is given by
\\
~\\
~\\
\unitlength 0.1in
\begin{picture}(34.10,11.35)(17.90,-19.35)
%
\special{pn 8}%
\special{pa 5200 800}%
\special{pa 2190 800}%
\special{fp}%
\special{sh 1}%
\special{pa 2190 800}%
\special{pa 2257 820}%
\special{pa 2243 800}%
\special{pa 2257 780}%
\special{pa 2190 800}%
\special{fp}%
\special{pa 2190 1600}%
\special{pa 5190 1600}%
\special{fp}%
\special{sh 1}%
\special{pa 5190 1600}%
\special{pa 5123 1580}%
\special{pa 5137 1600}%
\special{pa 5123 1620}%
\special{pa 5190 1600}%
\special{fp}%
%
\special{pn 8}%
\special{pa 5190 1200}%
\special{pa 2590 1210}%
\special{fp}%
\put(25.9000,-12.1000){\makebox(0,0)[lb]{$0$}}%
%
\special{pn 8}%
\special{ar 2190 1210 400 400  1.5707963 4.7123890}%
\put(33.9000,-20.2000){\makebox(0,0){{\bf Contour} $C$}}%
\end{picture}%

~\\

The contractions of the basic operators have
the following forms.
\begin{eqnarray}
U_j(\alpha_1)U_j(\alpha_2)=h_{jj}(\alpha_1-\alpha_2)
:U_j(\alpha_1)U_j(\alpha_2):~~\left
({\rm Im}(\alpha_2-\alpha_1)<
\frac{2\pi}{n}
,~j=0,n\right),\\
h_{00}(\alpha)=h_{nn}(\alpha)=e^{\gamma\frac{\xi-1}{\xi}
\frac{n-1}{n}}\frac
{
\Gamma_2(-i\alpha+\frac{2\pi}{n}\xi|\frac{2\pi}{n}\xi,2\pi)
\Gamma_2(-i\alpha+2\pi
|\frac{2\pi}{n}\xi,2\pi)
}
{\Gamma_2(-i\alpha+\frac{2\pi}{n}
|\frac{2\pi}{n}\xi,2\pi)
\Gamma_2(-i\alpha+2\pi+\frac{2\pi}{n}\xi-\frac{2\pi}{n}
|\frac{2\pi}{n}\xi,2\pi)
},
\end{eqnarray}

\begin{eqnarray}
U_j(\alpha_1)U_j(\alpha_2)&=&h_{jj}(\alpha_1-\alpha_2)
:U_j(\alpha_1)U_j(\alpha_2):~\left(
{\rm Im}(\alpha_2-\alpha_1)<\frac{2\pi}{n}(\xi-1),~
1\leq j \leq n-1\right),\\
h_{jj}(\alpha)&=&e^{2\gamma \frac{\xi-1}{\xi}}
\frac{\alpha}{i}\frac
{\Gamma_1(-i\alpha+\frac{2\pi}{n}\xi-\frac{2\pi}{n}
|\frac{2\pi}{n}\xi)}
{\Gamma_1(-i\alpha+\frac{2\pi}{n}|
\frac{2\pi}{n}\xi)},
\end{eqnarray}

\begin{eqnarray}
U_j(\alpha_1)U_{j-1}(\alpha_2)&=&h_{jj-1}(\alpha_1-\alpha_2)
:U_{j-1}(\alpha_1)U_j(\alpha_2):~\left(
{\rm Im}(\alpha_2-\alpha_1
)<\frac{\pi}{n},~1\leq j \leq n\right),\\
h_{jj-1}(\alpha)&=&e^{-\gamma \frac{\xi-1}{\xi}}
\frac{\Gamma_1(-i\alpha+\frac{\pi}{n}|\frac{2\pi}{n}\xi)}
{\Gamma_1(-i\alpha+\frac{\pi}{n}(2\xi-1)|\frac{2\pi}{n}\xi)},
\end{eqnarray}
and
\begin{eqnarray}
U_0(\alpha_1)U_{n}(\alpha_2)=h_{0n}(\alpha_1-\alpha_2)
:U_0(\alpha_1)U_n(\alpha_2):,~\left({\rm Im}(\alpha_2-\alpha_1)
<\frac{2\pi}{n}\right)\\
h_{0n}(\alpha)=e^{\frac{\gamma}{n}
\frac{\xi-1}{\xi}}\frac
{
\Gamma_2(-i\alpha+\pi+\frac{2\pi}{n}\xi-\frac{2\pi}{n}
|\frac{2\pi}{n}\xi,
2\pi)\Gamma_2(-i\alpha+\pi+\frac{2\pi}{n}|\frac{2\pi}{n}\xi,
2\pi)
}
{\Gamma_2(-i\alpha+\pi|\frac{2\pi}{n}\xi,
2\pi)
\Gamma_2(-i\alpha+\pi+\frac{2\pi}{n}\xi|\frac{2\pi}{n}\xi,
2\pi)
},
\end{eqnarray}

\end{appendix}

\end{document}